\RequirePackage{arydshln}
\documentclass[aps,twocolumn,nofootinbib,superscriptaddress,preprintnumbers,pra,10pt]{revtex4-1}

\usepackage{float}
\usepackage{colortbl}
\usepackage{amsmath,amssymb}
\usepackage{dsfont} 
\usepackage{hyperref}
\usepackage{graphicx}
\usepackage{enumitem}
\usepackage{arydshln}
\usepackage{mathtools}
\usepackage{bbold}
\usepackage{soul}
\makeatletter
\g@addto@macro\bfseries{\boldmath}
\makeatother

\newcommand{\be} {\begin{equation}}
\newcommand{\ee} {\end{equation}}
\newcommand{\bea} {\begin{eqnarray}}
\newcommand{\eea} {\end{eqnarray}}
\newcommand{\no} {\nonumber}
\newcommand{\cO}{{\mathcal O}}
\newcommand{\cL}{{\mathcal L}}
 
\def\bsll{{\ensuremath{b\to s \ell^{+}\ell^{-}}}\,}
\newcommand{\DCn}{\Delta \hat{C}^U_9}

\begin{document}


\title{On the significance of new physics in $b\to s\ell^+\ell^-$ decays}

\author{Gino Isidori}
\email{gino.isidori@physik.uzh.ch}
\affiliation{Physik-Institut, Universit\"at Zu\"rich, CH-8057 Z\"urich, Switzerland}
\author{Davide Lancierini}
\email{davide.lancierini@physik.uzh.ch}
\affiliation{Physik-Institut, Universit\"at Zu\"rich, CH-8057 Z\"urich, Switzerland}
\author{Patrick Owen}
\email{powen@physik.uzh.ch}
\affiliation{Physik-Institut, Universit\"at Zu\"rich, CH-8057 Z\"urich, Switzerland}
\author{Nicola Serra}
\email{nicola.serra@cern.ch}
\affiliation{Physik-Institut, Universit\"at Zu\"rich, CH-8057 Z\"urich, Switzerland}

\begin{abstract}
\vspace{5mm}

Motivated by deviations with respect to Standard Model predictions in \bsll decays, we evaluate the global significance of the new physics hypothesis in this system by including the {\it look-elsewhere effect} for the first time. 
We estimate the trial-factor with pseudo-experiments and find that it can be as large as seven. We calculate the global significance for the new physics hypothesis by considering the most general description of a non-standard \bsll amplitude of short-distance origin. Theoretical uncertainties are treated in a highly conservative way by absorbing the corresponding effects into a redefinition of the Standard Model amplitude.
Using the most recent measurements of LHCb, ATLAS and CMS, we obtain the global significance to be $4.3$ standard deviations.
\vspace{3mm}
\end{abstract}

\maketitle
\allowdisplaybreaks

\section{Introduction}\label{sec:intro}

Since 2013, several measurements have shown deviations from Standard Model (SM) predictions 
in rare $b$-hadron decays controlled by the underlying quark-level transition $b\to s\ell^+\ell^-$~($\ell=e,\mu$)~\cite{Aaij:2013qta,Aaij:2014ora,CMS:2014xfa,Aaij:2015oid,Aaij:2017vbb,Aaboud:2018mst,Sirunyan:2019xdu,Aaij:2019wad,Aaij:2020nrf,Aaij:2020ruw}.
The latest of these measurements provides the first evidence of a violation of Lepton Flavor Universality (LFU)
 in a single process~\cite{Aaij:2021vac}. 
 
 While there is no single result exhibiting a $5\,\sigma$ deviation from the SM,
the pattern of deviations, collectively denoted as the \bsll  {\em anomalies}, is striking. 
 In order to guide future activities in this field, and possibly claim a discovery,
it is essential to determine the combined statistical significance of these anomalies in a robust way. 
This is the purpose of this paper.

The first point to clarify is the alternative hypothesis that we aim to test with respect to the SM. The scope of this paper is to test 
in general terms the 
hypothesis of a new {\em short-distance} interaction connecting the $b$ and $s$ quarks with a dilepton pair. 
By short-distance we mean a NP interaction which appears as a local interaction in $b$-hadron decays. 
This general hypothesis, which is well justified by the absence of non-SM particles 
observed so far at colliders, allows us to describe \bsll transitions using the general formalism of effective Lagrangians, 
encoding a hypothetical NP contribution via appropriate four-fermion operators. 
This description, which is conceptually similar to Fermi's theory of beta decays~\cite{Fermi:1934hr}, allows to consider each specific 
$b$-hadron decay of interest as a different way to probe the same underlying $b\to s\ell^+\ell^-$  short-distance interaction.

The hypothesis of NP effects in \bsll transitions of short-distance origin
was formulated first in Ref.~\cite{Descotes-Genon:2013wba}. 
Later on several theory groups have analysed these processes within the framework 
of effective Lagrangians (see e.g.~Ref.~\cite{Altmannshofer:2013foa,Hurth:2013ssa,Hiller:2014yaa,Alonso:2014csa,Hurth:2014vma,Altmannshofer:2014rta,Descotes-Genon:2015uva,Ciuchini:2017mik,DAmico:2017mtc,Capdevila:2017bsm,Altmannshofer:2017fio,Alguero:2019ptt,Aebischer:2019mlg,Ciuchini:2020gvn}). 
These analyses provided  fits of the coefficients of well-defined sets of four-fermion operators, the so-called Wilson Coefficients (WCs), obtaining significances that in the last few years largely exceed the $5\,\sigma$ level~\cite{Alguero:2019ptt,Aebischer:2019mlg,Ciuchini:2020gvn}.  While these results are interesting and highly valuable, 
they do not provide the robust and general estimate of the significance we aim for.  
Our goal is not obtaining the best fit values of the WCs, which is the main goal of these previous studies, but rather estimating the significance of the NP hypothesis irrespective of its specific structure.

Most of the WC fits quoted in the literature are obtained by varying a small 
number of WCs, typically one or two. 
While this approach is well suited to test specific (often well motivated) NP hypotheses, and to determine the values of the WCs in these frameworks, it does not provide an unbiased estimate of the significance of the NP hypothesis. As we clarify below, the significance thus obtained 
resembles the {\em local significance} in resonance searches. 
The concern lies in the fact that several measurements are performed but only a few exhibit deviations with respect to the SM,
corresponding to well-defined sets of WCs.  It should also be stressed that the WC basis is a purely conventional choice: if a given correlation emerges from data in a two-parameter fit, one can change the basis and perform a fit with apparently higher significance enforcing such correlation via the basis choice and using a single parameter. 
  
 Overestimating the significance of a subset of measurements is equivalent 
to the {\it look-elsewhere effect} (LEE) in searches for new resonances~\cite{Lyons_2008,Gross:2010qma,Demortier:1099967}. 
While there is a small probability to observe a $n\sigma$ statistical fluctuation 
in a given bin of a distribution where the resonance could appear (local p-value), 
when several bins are measured the probability that at least one of them 
deviates by $n\sigma$ is larger (global p-value). 
When searching for a new resonance with unknown mass, the LEE can be addressed by calculating a trial-factor with an ensemble of pseudo-experiments~\cite{Demortier:1099967,Bock:353201,Ball:2007zza}, which is the ratio of the global and local p-values.
Conceptually, this is the same approach we adopt in this paper: We estimate the significance of the NP hypothesis in real data via pseudo-experiments. The trial-factor is then due to alternative deviations which could have emerged in a hypothetical dataset with the same experimental precision. 

There are fits in the literature that use a large number of WCs and a rather general NP hypothesis~\cite{Altmannshofer:2013foa,Alguero:2019ptt,Hurth:2021nsi}. In particular, Ref.~\cite{Hurth:2021nsi} fits all possible WC directions and therefore does not suffer from the LEE. The issue in this case is not the number of WCs but 
the effective number of degrees of freedom in the system, which depends on the correlations between WCs and the observables that are accessible to the experiments. Using pseudo-experiments is an efficient method to eliminate flat directions in the space of WCs and can easily account for many experimental details such as non-Gaussian uncertainties and correlated systematics. 

As far as theoretical uncertainties are concerned, 
the main concern are non-local contributions due to intermediate charm states. 
This subject has been widely discussed in the literature \cite{Lyon:2014hpa,Bharucha:2020eup,Gubernari:2020eft,Huber:2019iqf,Jager:2019bgk,Huber:2019iqf,Nakayama:2019eth,Arbey:2018ics,Chrzaszcz:2018yza,Blake:2017fyh,Ahmady:2015fha,Descotes-Genon:2015xqa,Ciuchini:2015qxb}. As a conservative choice, we simply disregard the extraction of  
short-distance information on amplitudes which might receive such non-local contributions.

Summarizing, the approach we propose to determine the statistical significance of NP in \mbox{\bsll} transitions
is based on the following points:
\begin{itemize}
    \item We consider the short-distance \mbox{\bsll} transition
     as a unique process constrained by different decay channels. 

    \item  We describe NP effects in \mbox{\bsll} transitions  
     using the most general effective Lagrangian compatible with the hypothesis 
     of an effective local interaction. 
    \item We estimate the trial-factor via an ensemble of pseudo-experiments generated according to the SM hypothesis and using the likelihood ratio as the test statistic.

    \item We adopt a highly-conservative attitude towards theory uncertainties, particularly in the case of non-local charm contributions.

\end{itemize}

This method allows us to evaluate the probability to observe the numerical coherence that is seen in data by chance. Only coherent deviations with respect to the SM can give a large value of the test statistic. All possible deviations in both the measurements and Wilson coefficients are considered. Therefore, this method evaluates the {\em global significance} of the 
\bsll  anomalies for the first time.

\section{Effective Lagrangian  and selection of the observables}\label{sec:WC}

In the limit where we assume no new particles below the electroweak scale, 
we can describe $b\to s\ell^+\ell^-$ transitions by means of an effective 
Lagrangian containing only light SM fields. The only difference between SM and 
effective Lagrangians, renormalized at a scale $\mu \sim m_b$, is the number of effective operators, 
which can be larger in the NP case.
To describe all the relevant non-standard local  contributions, we add to the SM 
effective Lagrangian
\be
\Delta \cL^{b\to s\ell\ell}_{\rm NP} = \frac{4 G_F}{\sqrt{2}}\,   
 \sum_{i } C_i \cO_i + {\rm h.c.}\,,
\ee 
where $G_F$ denotes the Fermi constant, and where the index $i$ indicates the following set 
of dimension-six operators (treated independently for $\ell=e$ and~$\mu$):
\bea
&& \cO^\ell_{9}=  (\bar{s}_L\gamma_\mu b_L)(\bar\ell \gamma^\mu\ell)  \,,  \quad\ 
\cO^\ell_{10}=  (\bar{s}_L\gamma_\mu b_L)(\bar\ell\gamma^\mu\gamma_5\ell) \, , \no \\
&& \cO^{\ell\prime}_{9}=  (\bar{s}_R\gamma_\mu b_R)(\bar\ell \gamma^\mu\ell)  \,,  \quad 
\cO^{\ell\prime}_{10}=  (\bar{s}_R\gamma_\mu b_R)(\bar\ell\gamma^\mu\gamma_5\ell) \, , \no \\
&& \cO^\ell_{\hat S}=  (\bar{s}_L  b_R)(\bar\ell_R \ell_L)  \,,  \qquad\
\cO^{\ell\prime}_{\hat S}=  (\bar{s}_R  b_L)(\bar\ell_L  \ell_R) \, .
\label{eq:Olist}
\eea
As shown in~\cite{Alonso:2015sja}, these operators are in one-to-one correspondence with the independent  
combinations of dimension-six operators involving $b$, $s$ and lepton fields in the complete  
basis of dimension-six operators invariant under the SM gauge group. 

We do not include in the list (\ref{eq:Olist}) the 
dipole operators, $\cO^{(\prime)}_7$, for two reasons: 
these do not describe a \bsll local interaction and
they are well constrained by $\Gamma(B\to X_s \gamma)$ and 
$\Gamma(B\to K^*\gamma)$.\footnote{An explicit quantification of the  change of the significance when $C^{(\prime)}_7$ are also varied, taking into account their {\em a priori}  knowledge before any LHCb measurements, is presented in Section~IV.}

The four scalar operators in (\ref{eq:Olist}) lead to \bsll  amplitudes which are helicity suppressed. 
We thus restrict the attention to the single effective combination which contributes to the $B_s^{0}\to \mu^+\mu^-$
helicity-suppressed rate. Finally, in the absence of stringent 
experimental constraints on CP-violating observables, we treat the NP WCs as real parameters.\footnote{This statement 
refers to the standard quark-phase convention, where the WCs are approximately real also in the SM. 
Imaginary contributions to the WCs would not interfere with the SM amplitude and cannot induce large 
deviations from the SM in CP-conserving observables.   }
According to these general hypotheses, NP effects in \bsll  transitions
are described in full generality by nine real parameters. As far as $C^{e,\mu}_{9,10}$
are concerned, it is convenient to separate universal and non-universal corrections in lepton flavor, defining
\begin{align}
\begin{aligned}
& C_i^e= C_{i}^\mathrm{SM} + \Delta C_i^U\,,   \\ 
& C^\mu_i = C_{i}^\mathrm{SM}  +\Delta C_i^U + \Delta C_i^{\mu }  \,. 
\label{eq:C9C10mue} 
\end{aligned} 
\end{align}

Adopting a conservative attitude toward theoretical uncertainties, 
we restrict the attention to the following three sets of observables:
i)~the LFU ratios $R_K$~\cite{Aaij:2021vac}  and $R_{K^*}$~\cite{Aaij:2017vbb}, ii)~the branching ratio for the rare dilepton mode $B_s^{0}\to \mu^+\mu^-$~\cite{CMS:2014xfa,Aaboud:2018mst,Sirunyan:2019xdu,Beneke:2019slt} and, iii)~the normalized angular distribution in $B^{0}\to K^{*0}\mu^+\mu^-$ decays~\cite{Aaij:2020nrf,Aaij:2020ruw}. As the measurements in class
i) and ii) are statistically dominated, they are treated as uncorrelated whereas the full experimental correlation matrix given in Ref~\cite{Aaij:2020nrf} is used for the $B\to K^{*}\mu^+\mu^-$ angular observables.

By construction, the observables in class
i) and ii) are insensitive to form-factor and decay constant uncertainties (except for $f_{B_s^{0}}$ in class~ii) as well as non-local charm contributions. 
The latter
induce contributions to the decay amplitudes that 
can effectively be described via the shift 
\be
 \Delta C_9^U \to  \Delta C_9^U + f^{c\bar c}_{B \to f} (q^2) 
\ee
where $q^2$ denotes the squared dilepton invariant mass. 
The absence of a completely reliable estimate of the theoretical uncertainty on the function $f^{c\bar c}_{B \to f}$, 
in particular on its normalization at $q^2=0$, forces us to treat the determination of  
$\Delta C_9^U$ as SM nuisance parameter\footnote{If we were to include more channels potentially affected by
non-local charm contributions, we would need to treat the determination of 
$\Delta C_9^U$ from  each channel as an independent nuisance parameter.}
and ignore the information from exclusive decay rates or dilepton spectra.
This way we automatically remove of most of the uncertainties associated to the hadronic form factors:  a choice that maybe seen as too conservative, but that certainly does not lead to overestimate the NP significance. 

The only observable with a residual 
form-factor uncertainty we retain is 
the $B^{0} \to K^{*0}\mu^+\mu^-$ 
angular distribution. We keep it since 
this distribution is sensitive to 
non-standard effects in short-distance operators other than 
$\cO^\mu_9$, even if we marginalise over $\Delta C_9^U$. To reduce the form-factor uncertainty we make use of the $P_i$ observables~\cite{Matias:2012xw}. 
We explicitly checked that consistent results are obtained using 
the $S_i$ observables~\cite{Altmannshofer:2008dz},
employing the form-factor parameterization in~\cite{Straub:2015ica}.

The set of nine parameters discussed above provides an unbiased description of heavy 
NP contributions to 
{\mbox \bsll} transitions. In order to evaluate the impact of motivated, but 
more specific theoretical assumptions, we also define a reduced set of WCs based 
on the hypotheses of small flavor-violating effects in the right-handed sector. According to this 
hypothesis, $C^{\ell\prime}_i \approx0$ and the set of independent WCs is reduced to five operators.
This hypothesis follows from the general assumption of a minimally broken $U(2)^3$ flavor 
symmetry: A general property of SM extensions which was proposed in~\cite{Barbieri:2011ci} well before the observation 
of the \bsll anomalies,
motivated by the  stringent constraints on right-handed quark flavor mixing especially in the kaon system (see 
e.g.~\cite{Davidson:2007si}).

\section{Statistical Method}\label{sec:Stat}
To evaluate the significance of the NP hypothesis in the $b\to s\ell^+\ell^-$ system we use
\be
\Delta \chi^2 =-2\log\frac{{\cal L}(X|\DCn , C_i^{\rm SM})}{{\cal L}(X|\hat{C}_i)}
\ee
as the test statistic. The trial-factor is calculated with a similar technique as described in Ref.~\cite{Demortier:1099967,Bock:353201,Ball:2007zza}. Starting from SM predictions, a large number of pseudo-experiments are generated, varying the measurements according to the experimental uncertainty. For each simulated experiment, the full set of WCs ($C_i$) is fitted and the $\Delta \chi^2$ between the best fit ($\hat{C}_i$) and the SM prediction ($\DCn$, $C_i^{\rm SM}$) is calculated. 
Data are fitted in the same way as pseudo-experiments and the distribution of $\Delta \chi^2$ is used to calculate the p-value. 
The software package {\it Flavio}~\cite{Straub:2018kue} is used to fit WCs. 

One of the interesting features of the \mbox{\bsll} anomalies is that they can be easily explained with only one WC: $C_{LL}^{\mu}=\Delta C_9^{\mu}- \Delta C_{10}^{\mu}$. While this makes the NP hypothesis easy to interpret from the theory point of view, it is not the best way to assess the sensitivity with respect to the SM. To illustrate this point we apply our method to the fit of one or two WCs. 
\begin{figure}[t]
\centering
\includegraphics[width=0.40\textwidth]{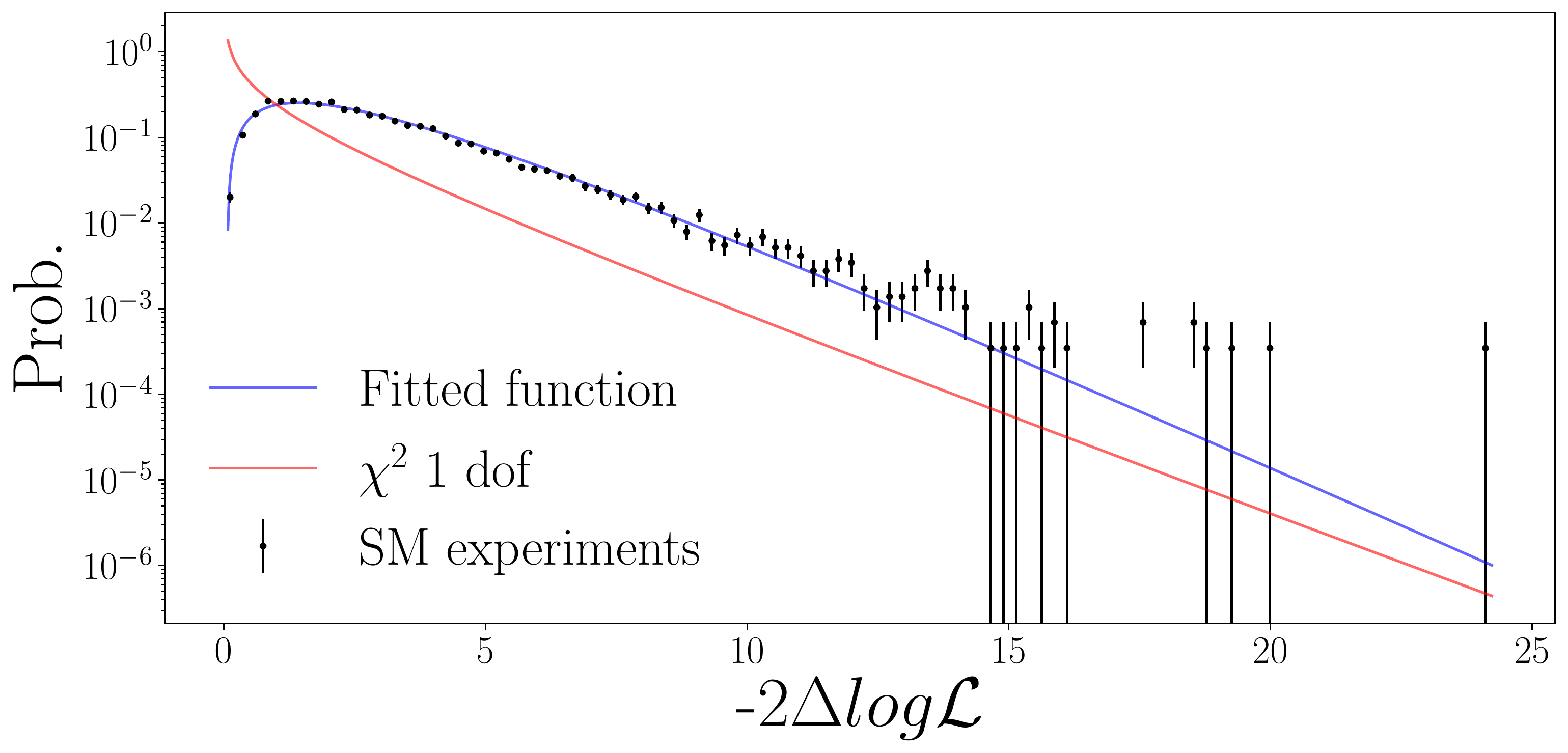}
\includegraphics[width=0.40\textwidth]{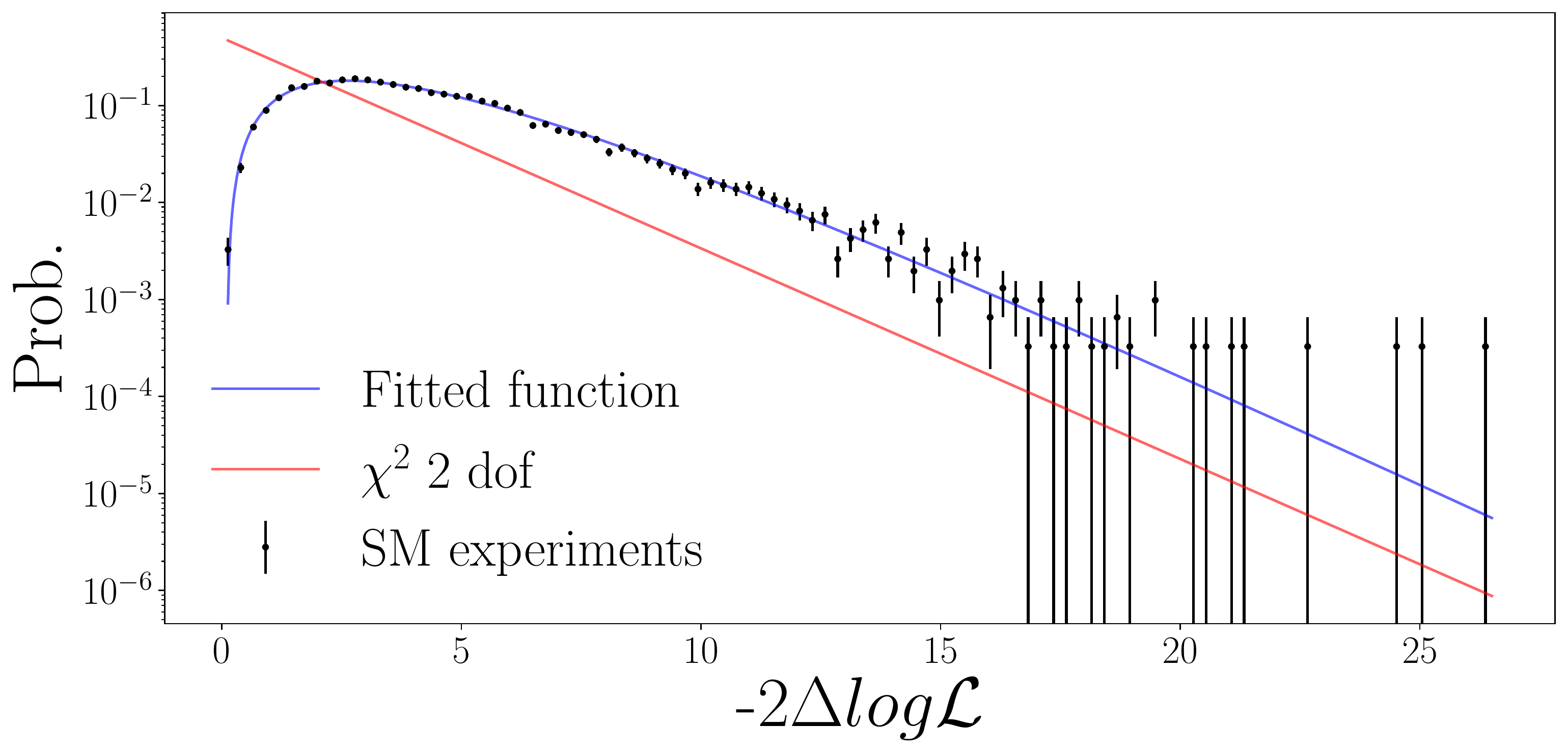}
\caption{$\Delta \chi^2$ distribution
extracted from pseudo-experiments
(blue) for fitting the best one/two WCs varying the SM, compared to the theoretical
$\chi^2$ distribution with one/two degrees of freedom (red).}
\label{fig:DeltaChi2_WC1,2}
\end{figure}
In Fig.~\ref{fig:DeltaChi2_WC1,2}, the $\Delta \chi^2$ distribution under the SM hypothesis is shown when the one/two WCs which maximise the likelihood are chosen to fit the data: For each pseudo-experiment, we fit every single possible one/two WC combination and choose the largest test statistic. The blue curve is an empirical function that best describes the distribution. The comparison with a $\chi^2$ distribution with one/two degrees of freedom demonstrates that a sizeable trial-factor is present. 
Taking for instance a hypothetical $4\,\sigma$ discrepancy when fitting the best one/two WCs, it would be diluted down to 3.7/3.5$\,\sigma$ with a trial-factor equal to 4.1/7.0, respectively. 
Since the current best scenarios to explain the anomalies with NP in $C_{LL}^{\ell}$ or in $C_9$ and $C_{10}$ have emerged from the data, using this hypotheses to evaluate the NP significance can lead to overestimates. 


As discussed in Sec.~\ref{sec:WC}, we advocate the full set of nine WCs to be used if we would like to have an agnostic approach to NP.   
However, the full set of WCs contains redundancy, which makes the fit unstable. 
For instance, the deviations in $R_K$ and $R_K^*$ can be explained with non zero values of $C_{LL}^{\mu}$ or non-zero values of $C_{RL}^\mu=C_{9}^{\mu\prime}-C_{10}^{\mu\prime}$. Here we are not interested in interpreting the best NP direction and we therefore treat all of these in the same way. In total, the maximum number of WCs that can be fitted is seven, with the full basis of muonic operators, the single effective combination of scalar operators, and two electronic operators.
Each pseudo-experiment is fitted six times, with all possible combinations of seven WCs. For each experiment, the largest test-statistic value is used. Adding redundant directions will not improve the $\chi^2$ of a given pseudo-experiment, since there are not enough sensitive measurements to constrain simultaneously all nine WCs.

\section{Results}\label{sec:Res}
The $\Delta \chi^2$ distribution for the fit to the full set of Wilson coefficients is shown in Fig.~\ref{fig:DeltaChi2_WC9,5} (top).  
The same procedure is then used in data, obtaining a $\Delta \chi^2=31.4$, which corresponds to a global significance of $4.3\sigma$. As expected, the large $\Delta \chi$ value arises mostly due to the discrepancies with respect to the SM in the LFU ratios, $R_K$ and $R_K^{*}$. 
\begin{figure}[t]
\centering
\includegraphics[width=0.40\textwidth]{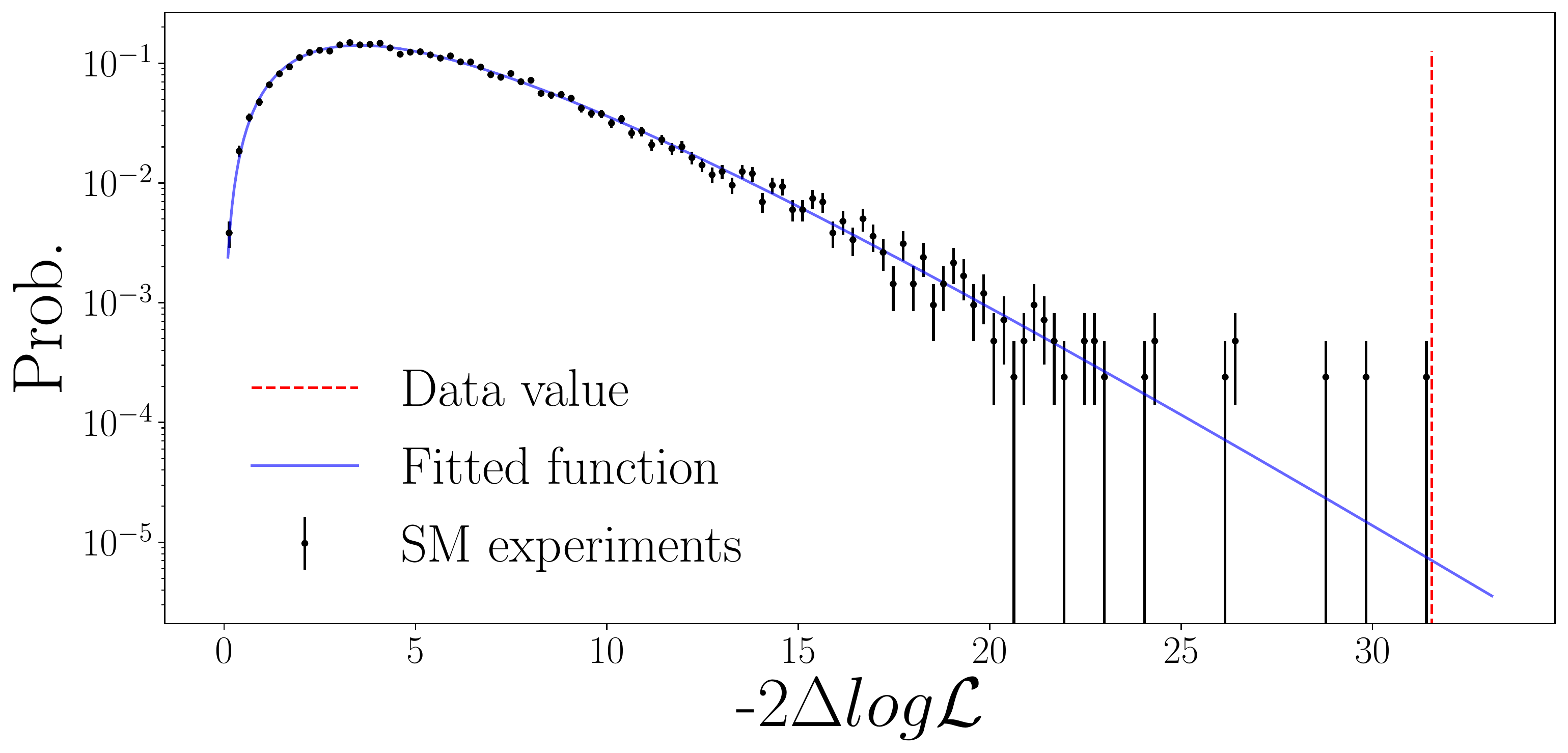}
\includegraphics[width=0.40\textwidth]{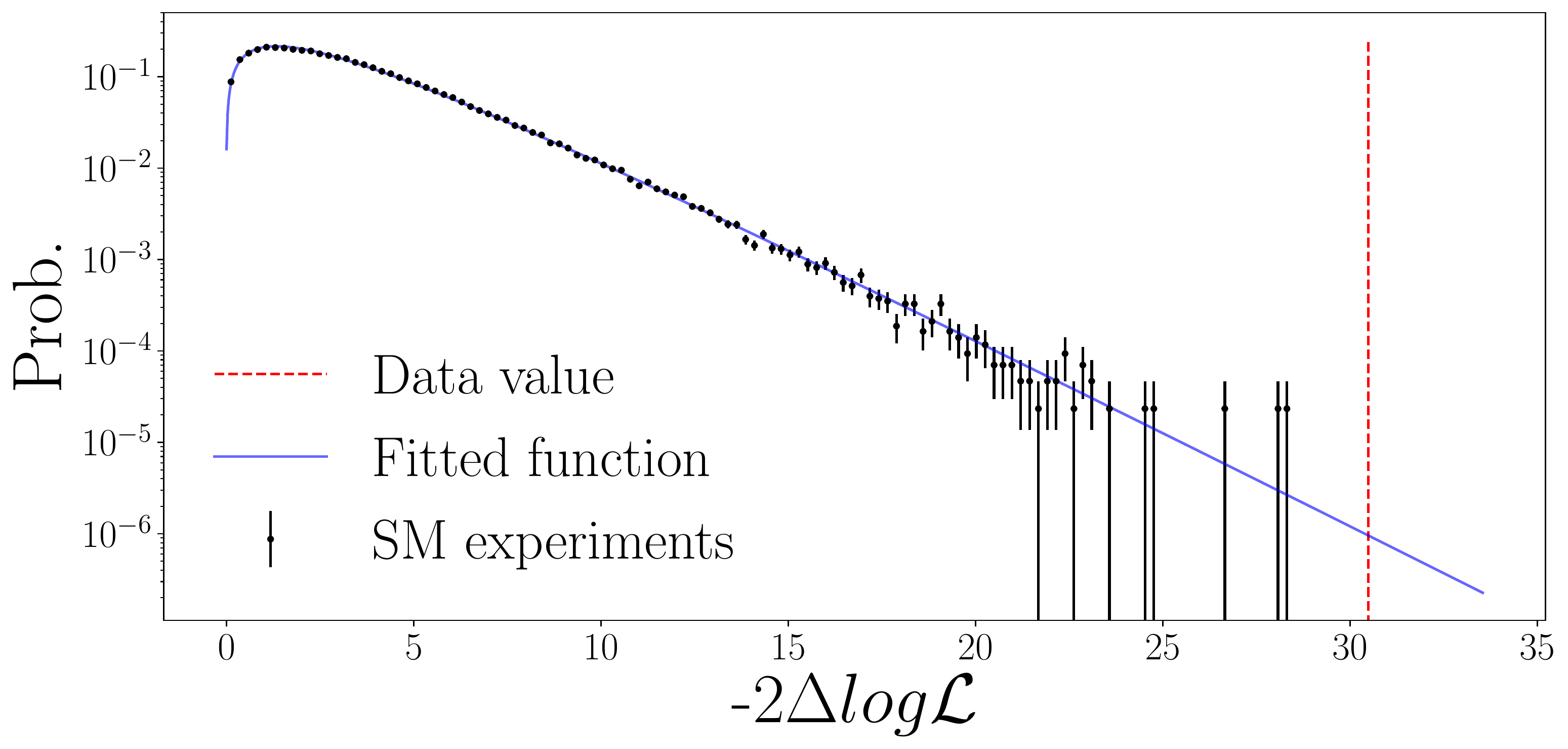}
\caption{$\Delta \chi^2$ distribution (blue) for SM pseudo-experiments in the general 9 WC fit basis (top) and the reduced 5 WC basis (bottom). The data is shown as a vertical red line on the plot.}
\label{fig:DeltaChi2_WC9,5}
\end{figure}
The goodness of fit to data can be computed by calculating the p-value of the absolute $\chi^2$ of the best fit. This results in a $11\%$ p-value, which is acceptable. The largest pulls of the best fit with respect to the measurements come from the lowest $q^2$ bins of the angular observables in the $B^0\to K^{*0}\mu^{+}\mu^{-}$ decays. This is a known issue~\cite{Alguero:2019pjc} and has a small impact on the significance. Eliminating the lowest $q^2$ bin of all the angular observables decreases the $\Delta \chi^2$ by only one unit and the fit quality of the fit improves, leading to a p-value associated to the absolute $\chi^2$ of $24\%$.

While the $C_7^{(\prime)}$ WCs do not describe $b\to s\ell\ell$ contact interactions and are not included in the default analysis, we investigated the impact of adding them to the set of WCs we allow to be 
affected by NP. 
Imposing constraints on $C_7^{(\prime)}$ prior to the flavour anomalies from Ref~\cite{HFLA_bsgamma} and including the angular analysis of $B^0\to K^* e^+e^-$ from Ref~\cite{LHCb:2020dof}, the total significance marginally decreases, as expected, from 
$4.3\sigma$ to $4.2\sigma$.

Here we advocate that for claiming a discovery, the NP significance should be calculated using an agnostic approach. However, as discussed in Sec.~\ref{sec:WC}, there were good \emph{a-priori} theoretical reasons to assume no NP in $C^{\ell\prime}_{9,10}$. To evaluate the significance of this hypothesis we apply our method to the reduced set of five WCs. The $\Delta \chi^2$ distribution is shown in Fig.~\ref{fig:DeltaChi2_WC9,5} (bottom)\footnote{The  5 WCs fit has the same goodness of fit as the 9 WCs fit, since data can be well described with a smaller number of Wilson Coefficients.}. 
Applying the same fit to data we obtain a $\Delta \chi^2=30.5$, which integrating the distribution corresponds to a significance of $4.7\sigma$.
Interestingly, this is similar to the values quoted in the recent literature~\cite{Geng:2021nhg,Altmannshofer:2021qrr,Cornella:2021sby} for single-parameter fits of theoretically clean observables only.
Having a larger number of free parameters, one could have expected a lower significance in our case. However, in this specific case 
the LEE effect is compensated by two 
facts: i)~the inclusion of the angular distribution of the $B\to K^{*}\mu^+\mu^-$ decay which, even after marginalizing  over $\Delta C_9^U$, retains some sensitivity 
to the other WCs; ii)~the overall higher $\Delta \chi^2$ obtained with more parameters. This observation reinforces the high significance of 
the \bsll anomalies in motivated NP models.

\section{Conclusion and discussion}

In conclusion, we have presented a method to evaluate the global significance for the NP interpretation of the \bsll anomalies. This method transposes the known criteria used for discovering new resonances, such as the Higgs boson, into searching for NP in $b\to s\ell^+\ell^-$ transitions. 
It is worth emphasizing that, while it is remarkable that all data can be explained by fitting one or two WCs and that this observation can be used to investigate what are the interesting theoretical directions, this hypothesis has been made after having seen the data. 
Using the same hypothesis to evaluate the global significance of NP would be the Bayesian-inference equivalent of choosing the prior after having calculated the likelihood. 
Therefore, we advocate a more agnostic method to calculate the global NP significance with respect to the SM in $b\to s\ell^+\ell^-$ processes. 
To this end, we have calculated the LEE for the first time and shown that the trial-factor cannot be neglected. 


We stress that the approach proposed in this paper should not be interpreted as a criticism towards existing attempts
made so far of combining and interpreting the anomalies in motivated theoretical frameworks.  We are simply addressing a different question. 
While current fits of selected WC  sets in the \bsll system only evaluate a local significance, 
these approaches are fundamental to obtain theory insights on the flavor anomalies. 
Similarly, there is a strong theoretical interest in trying to combine the \bsll  anomalies with other hints of 
deviations from the SM,  such as the $b\to c\ell\nu$ anomalies~\cite{Lees:2012xj,Aaij:2017deq,Lees:2013uzd,Sato:2016svk,Aaij:2015yra,Huschle:2015rga,Aaij:2017tyk,Belle:2019rba,Hirose:2017dxl} or the recent 
 $(g-2)_\mu$ result~\cite{Abi:2021gix,Albahri:2021ixb} (see also~\cite{Borsanyi:2020mff}).
 However, this combination is not appropriate to establish a global significance, 
 given  the hypothesis of a connection between different processes 
 is made a posteriori, after having observed data.

We also recognise that our approach of treating $\Delta C_9^U$ as a nuisance SM parameter can be viewed as an overly conservative choice.
Nevertheless, in the absence of a widely accepted estimate for the theory uncertainty of the non-local $c\overline{c}$ contributions, 
this is mandatory for a conservative estimate of the significance. 

While the uncertainty of all the measurements used here are statistically dominated, the results of our analysis can be improved by adding  correlations of experimental systematic uncertainties and taking into account that they can follow non-Gaussian PDFs. Additional potential 
improvements concern the observables to be included.
To simplify the numerical analysis we have only included the observables that are most sensitive.

For instance, observables such as $Q_5$~\cite{Alguero:2019pjc} measured by Belle~\cite{Wehle:2016yoi}, were not included in this work since these measurements are still not precise enough to have a sizeable impact.
For the same reason, angular observables in $B_s^{0}\to \phi \mu^+\mu^-$~\cite{LHCb:2021xxq,LHCb:2015wdu} decays are not considered. 
While the decay $B_s^{0}\to \phi \mu^+\mu^-$ is analogous to $B^0\to K^{*0}\mu^+\mu^-$ from theory point of view, it is limited statistically due to the value of the fragmentation fraction $f_s/f_d$~\cite{LHCb:2021qbv} and that it is not self-tagged decay.

While beyond the scope of this paper, a more rigorous approach of including all observables and treat all correlated systematics 
is desirable in view of future combinations.

With current data, all these effects are expected to have a small impact 
and will not change the main conclusions presented here.

The global significance of $4.3$ standard deviations we obtain for the NP hypothesis in the \bsll system clearly demonstrates the potential of combining different measurements in this system, even when adopting an agnostic alternative hypothesis and an highly conservative theory approach. 
In view of future measurements,  we advocate that experimental collaborations 
adopt this method to calculate the global significance of the new physics hypothesis in a conservative and unbiased way.


\acknowledgements

This work was inspired by a discussion with Niels Tuning and a separate discussion with Joaquim Matias. We acknowledge their contributions in attracting our attention to this problem. We thank Konstantinos Petridis 
and Diego Tonelli for very useful discussions and comments. We also thank Abhijit Mathad for valuable crosschecks on the numerical results of the paper.
This project has received funding from the Swiss National Science Foundation (SNF) under contracts 00021-182622 and 200021-175940, and from
the European Research Council (ERC) via the European Union's Horizon 2020 research and innovation programme under grant agreement 833280 (FLAY).

\bibliography{letter_PLB}

\end{document}